\documentclass[aps,twocolumn,showpacs,amsmath,amssymb,pre,superscriptaddress, floatfix,nofootinbib]{revtex4-1}

\usepackage{graphicx}
\usepackage{dcolumn}
\usepackage{bm}
\usepackage{amssymb}
\usepackage{multirow}
\usepackage{color}

\newcommand{\ea}{\begin{eqnarray}}
\newcommand{\ee}{\end{eqnarray}}
\newcommand{\4}[2]{{\frac{#1}{#2}}}
\newcommand{\Sum}[2]{{\sum\limits_{#1}^{#2}}}

\begin{document}
\title{Interaction-induced current-reversals in driven lattices}
\date{\today}
\author{Benno Liebchen}
\email[]{Benno.Liebchen@physnet.uni-hamburg.de}
\affiliation{Zentrum f\"ur Optische Quantentechnologien, Universit\"at Hamburg, Luruper Chaussee 149, 22761 Hamburg, Germany}%
\author{Fotis K. Diakonos}
\affiliation{Department of Physics, University of Athens, GR-15771 Athens, Greece}%
\author{Peter Schmelcher}
\email[]{Peter.Schmelcher@physnet.uni-hamburg.de}
\affiliation{Zentrum f\"ur Optische Quantentechnologien, Universit\"at Hamburg, Luruper Chaussee 149, 22761 Hamburg, Germany}%

\begin{abstract}
We demonstrate that long-range interactions can cause, as time evolves, consecutive reversals of directed currents for dilute ensembles of particles in driven lattices.
These current-reversals are based on a general mechanism which leads to an interaction-induced accumulation of particles in the regular regions of the underlying
single-particle phase space and to a
synchronized single-particle motion as well as an enhanced efficiency of Hamiltonian ratchets.
\end{abstract}
\maketitle

\begin{figure}[htb]
\begin{center}
\includegraphics[width=0.48\textwidth]{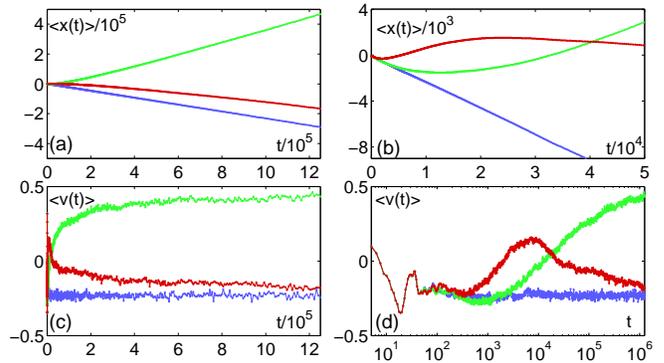}
\caption{Time-evolution of the ensemble average of the position $\langle x\rangle(t)$ (a), its magnification for short and intermediate times (b), the average velocity $\langle v\rangle(t)$ (c) also with semilogarithmic
axis (d). Parameters: $a=\omega=V=\beta=1; L=10$; $\alpha=0$ (blue/medium saturation); $\alpha=0.8$ (green/low saturation) and $\alpha=8.0$ (red/full saturation).}
\label{means}
\end{center}
\end{figure}

\begin{figure}[htb]
\begin{center}
\includegraphics[width=0.4\textwidth]{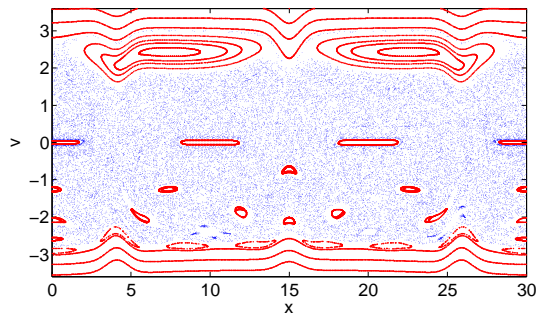}
\caption{Stroboscopic Poincar\'{e} surface of section of the single-particle phase space ($N=1$) at times $t/T \in \mathbb{N}$.
Parameters: $a=\omega=V=\beta=1; L=10$. Blue dots: chaotic sea. Red lines: regular orbits.}
\label{psos}
\end{center}
\end{figure}

\begin{figure}[htb]
\begin{center}
\includegraphics[width=0.48\textwidth]{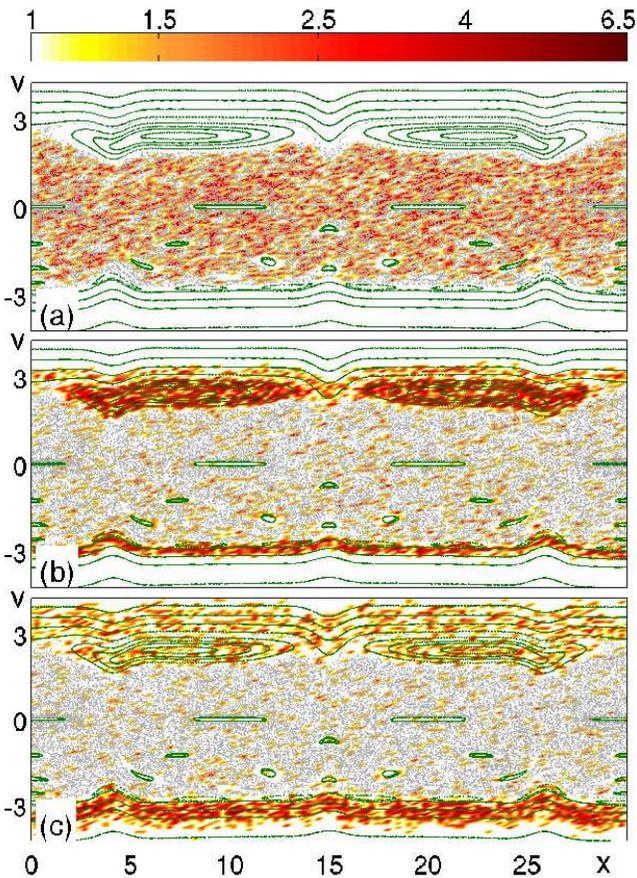}
\caption{Superposition of the single-particle Poincar\'{e} surface of section (grey/green) and the distribution of particles in the phase space at $t=2\cdot 10^5 T$ (red/yellow).
Parameters: $a=\omega=V=\beta=1; L=10$; $\alpha=0$ (a); $\alpha=0.8$ (b) and $\alpha=8.0$ (c).
}
\label{overlap}
\end{center}
\end{figure}

\paragraph*{Introduction}
Ratchets involve nonequilibrium dynamical processes which make it possible to realize the old desire of converting thermal fluctuations into directed currents (DCs) and therefore usable work.
According to their generality and 
their relevance for devices like Brownian \cite{astumian97,astumian02,reimann02} and molecular motors \cite{julicher97} but also for
biophysical problems like the migration of bacteria \cite{lambert10} or the cell mobility in cancer metastasis \cite{mahmud09}, 
the need for an understanding of the detailed properties and prospects of the mechanisms of ratchets has
given rise to the development of a highly active research field. This ranges from 
the theoretical analysis of the generating mechanism of DCs  \cite{magnasco93,flach00,schanz01,denisov07}
to the remarkable realizations of ratchets in setups as different as semiconductor nanostructures \cite{linke99}, Josephson junction arrays \cite{sterck02} and 
optical lattices \cite{schiavoni03,gommers05,salger09} (see also the review \cite{renzoni11}).
Recently, it has been demonstrated that lattices with a spatially dependent driving imply a tunable phase space \cite{petri10} and enrich 
the physics of DCs with mechanisms allowing for the creation of traveling density waves 
\cite{petri11} and designable patterned particle deposition \cite{liebchen11}.
All of the above investigations do not focus on particle interactions and indeed most works on interacting ratchets concentrate
 either on the stochastic or the overdamped deterministic case \cite{wambaugh99,denisov05,chepelianskii08,ai11}
leaving a gap in the literature concerning the microscopic analysis of interacting (deterministic) ratchets.
A topic of particular interest are the so-called current-reversals \cite{jung96,schreier98,linke99,mateos00,silva06,marconi07,petri10}, i.e. the 
tunability of the orientation of DCs via system parameters or particle density. 
Very recently even a current-reversal occurring in the time-evolution
was achieved by a time-dependent modulation of the asymmetry of the ratchet potential \cite{arzola11}.
\\In the present work we demonstrate 
 that long-range interactions in a dilute particle ensemble cause self-driven current-reversals without 
requiring time-dependent modulations of external parameters. 
We analyze this surprising phenomenon to be the expression of a more general mechanism based on the interplay of two-body collisions and the
underlying driven single-particle dynamics. 
As a consequence, we obtain synchronized single-particle motion as well as an increase of DCs.
\paragraph*{Setup}
We consider a system of $N$ equally charged particles in a one-dimensional lattice of laterally oscillating Gaussian potential barriers with amplitude $a$, frequency $\omega$, height $V$ 
and equilibrium distance $L$ described by the Hamiltonian
\ea \label{hamilt} H &=& \Sum{i=1}{N} \left[
\4{p_i^2}{2m} + V \Sum{j=-\infty}{\infty}{\rm e}^{-\beta\left[x_i-a f_j(t) - j \cdot L \right]^2} \right. \nonumber \\ &+& \left. 
\Sum{k=1}{i-1}\4{\alpha}{\left|x_i - x_k \right|}  \right]; \quad f_j(t)=\cos(\omega t + \phi_j)
\ee
Therein $x_i,p_i,m$ denote the position, momentum and mass of the $i$-th particle,
$f_j(t)$ is the driving law of the barrier with equilibrium position $j\cdot L$ and 
$\alpha:=q^2/4\pi \epsilon$ represents the interaction coefficient, where $q$ is the charge of the particles and $\epsilon$ is the permittivity.
After performing the scaling transformations $(x,t,m)\mapsto (x':=x/a, t':=t \omega, m'=m \omega^2 a^2/V)$, we may set $a=\omega=V=1$ without loss of generality. 
For the remaining parameters we choose $m=\beta=1$; $L=10$ and simulate 125 ensembles each consisting of $N=80$ interacting particles by numerical integration of the corresponding Hamiltonian equations of motion.
Large scale numerical computations have been performed for the integration of the resulting 160 coupled, nonlinear ordinary differential equations 
for $t/T=2\cdot 10^5$ periods of the driving law.
Even though our results are essentially independent of the initial conditions, our focus is on dilute ensembles whose initially stored interaction
energy and kinetic energy is small in comparison to $V$
(for simplicity all particles are placed equidistantly with a spacing $L_0:=x_{i+1}-x_{i}$). 
The total energy concerning the interactions of the $i$-th particle with the others 
reads $V_{\rm ia}(x_i)=(\alpha/L_0)[H_{i-1}+H_{N-i}]$ where $H_k=\Sum{i=1}{k}(1/j)$ is the $k$-th harmonic number. 
Choosing $L_0=15L$ and $\alpha \leq 8.0$ we have $V_{\rm ia}(x_i) \lesssim 0.06 \alpha < V$.
The initial particle velocities are randomly chosen in the low velocity region of the chaotic sea of the corresponding single-particle phase space (SPPS) resulting from Eq.~\ref{hamilt} for $N=1$.
For $f_j(t)$, we choose a harmonic driving law and in order to break the symmetries necessary to obtain a DC (\cite{flach00})
phase shifts of period three $(\phi_1,\phi_2,\phi_3)=(0,2\pi/3,4\pi/3);\;\; \phi_{i+3}=\phi_i$ are applied \cite{petri10}.
It is important to note that the observed effects and the underlying mechanism are generally valid and therefore not restricted to a specific driving as long as the
relevant symmetries are broken. 
\paragraph*{Current-reversals and fast ratchets}
We now explore the impact of long-range interactions on the directed transport.
Fig.~\ref{means} (a) - (d) show the time evolution of the mean position and velocity of the ensemble of particles for three different interaction strengths 
which can be adjusted according to the above-mentioned scaling behaviour of the system.
In Fig.~\ref{means} (a) which shows the long-time behaviour for $\alpha=0$ we observe a directed particle current with a negative transport velocity $v_T \sim -0.28$.
In the presence of interactions, we first note, that the directed transport persists.
Surprisingly, for the interaction strength $\alpha=0.8$ the direction of transport is inverted.
For $\alpha=8.0$, the DC points again in the same direction as for the noninteracting case.
Fig.~\ref{means} (b) shows that in the noninteracting case
a constant transport velocity is reached already after a short time $t \sim 10^2$. 
In contrast to this, for $\alpha=0.8$ the DC initially points in the same direction as for $\alpha=0$ and possesses a similar velocity for several hundred of periods of the driving law but
for longer times the particle current continuously slows down and reverts its direction.
Even more, we observe a magnitude of the reverted transport velocity which is larger than for the noninteracting case (Fig.~\ref{means} (c)).
Thus, we encounter two phenomena: An interaction induced current-reversal and an increase of the ratchet-efficiency ('fast ratchet').
For $\alpha=8.0$ even two current-reversals can be observed (Fig.\ref{means} (a,b)). 
The DC deviates earlier from its noninteracting counterpart than for $\alpha=0.8$, but inverts its direction for a second time. 
Most clearly, the situation is reflected in Fig.~\ref{means} (d):
Due to their small initial velocities the particles remain for the first few collisions on their lattice sites and are shaken according to the force provided by the driving law, so that their mean velocity simply follows the latter.
Then, for $\alpha=0.8$ and $\alpha=8.0$ the mean velocity increasingly deviates from its noninteracting counterpart. 
We remark that in spite of the significant computational effort leading to the results shown in Fig.\ref{means} the asymptotic behaviour for long times is not yet reached. 
Indeed, below we will argue why the asymptotic states are reached only for such long times.
\paragraph*{Particle accumulation mechanism}
We derive now the mechanism which is responsible for the current-reversals.
The backbone of our analysis is the single-particle dynamics in the absence of all interactions. 
Since our many-particle system is dilute, the impact of the Coulomb interaction on the dynamics represents, most of the time, a small perturbation 
of the interaction-free dynamics. However, as we shall see there are events, namely two body collisions, for which the Coulomb interaction plays a significant role and which 
are responsible for the observed phenomena.
Let us therefore first analyze the structure of the SPPS via 
the stroboscopic single-particle Poincar\'{e} surface of section (PSOS) shown in Fig.~\ref{psos}. 
The phase space is mixed, i.e. we encounter a large chaotic sea at low velocities with embedded elliptic islands and invariant spanning curves confining the chaotic sea.
The elliptic islands correspond to synchronized particle-barrier motion in configuration space.
Islands crossing the $v=0$ line belong to particles which are trapped between two of the Gaussian potential barriers, whereas the others 
belong to particles ballistically moving in a given spatial direction. 
The first invariant spanning curve (FISC) limits the energy which is maximally achievable for chaotic trajectories. 
Motion on these invariant spanning curves involves energies which allow the ballistically flying particles to traverse the barriers for any phase of their motion. 
We observe an apparent asymmetry w.r.t. $v=0$ in the PSOS (Fig.~\ref{psos}), which reflects the breaking of the relevant symmetries \cite{flach00} and is responsible for the occurrence of directed transport in the noninteracting system. 
\\For the interacting many-particle system it is not possible to visualize the underlying high-dimensional phase space and 
we focus on specific observables to analyze and understand the time-evolution. Let us first inspect the time-evolution of the velocity distributions.
\begin{figure}[htb]
\begin{center}
\includegraphics[width=0.48\textwidth]{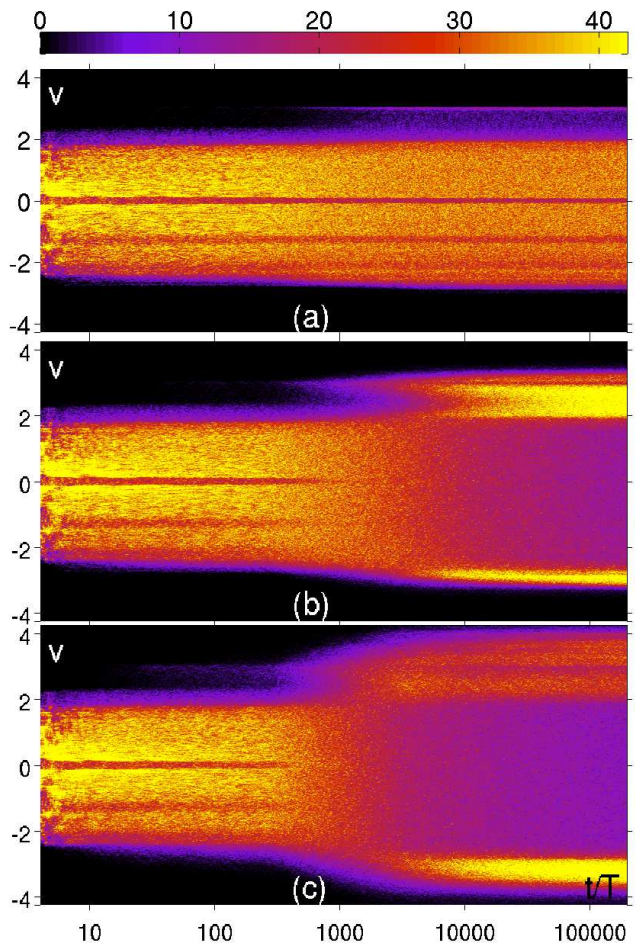}
\caption{Time-evolution of the velocity distribution for $125$ ensembles of $80$ particles. The number of particles (color) is shown as a function of time (logarithmically) and velocity. 
(a) $\alpha=0$; (b) $\alpha=0.8$; (c) $\alpha=8.0$. Ensemble and parameters as in Fig.~\ref{means}.}
\label{tev}
\end{center}
\end{figure}
For $\alpha=0$ (Fig.~\ref{tev} (a)) we encounter a uniform distribution in regions where we have no relevant elliptic islands
in the phase space (Fig.~\ref{psos}) and dips at velocities where large islands are located.
The sharp decrease at $v \sim 3.0$ and $v \sim - 2.7$, for times $t > 100$ when the single-particle phase space is occupied uniformly, reflect the velocity-boarders of the FISCs. 
The occurrence of a directed transport with negative velocity is predominantly due to the strongly reduced density for velocities $v \gtrsim 1.9$ corresponding to the large elliptic islands
centered at $v \sim 2.5$ (Fig.~\ref{psos}). 
\\For $\alpha=0.8$ (Fig.~\ref{tev} (b)) and for short times, the velocity distribution is similar to the corresponding distribution for $\alpha=0$, reflecting the observed equality of the transport velocity in both cases. 
Subsequently, the fine structure of the SPPS, i.e. the dips in the velocity-distribution, dies out continuously.
Simultaneously the particles start to accumulate at velocities $v \sim 2-3$ and around $v \sim -2.5$. 
Since the accumulation of particles around $v \sim 2.5$ is dominant for large times, we obtain a DC pointing in the opposite direction as for $\alpha=0$, i.e. a 
current-reversal is encountered.
For $\alpha=8.0$ (Fig.~\ref{tev} (c)) a faster broadening of the velocity-distribution compared to $\alpha=0.8$ is observed.
From $t \sim 200$ on a particle accumulation develops in the velocity region $v \sim 2-3$ (first current-reversal) and spreads out up to $v \sim 3.9$. Subsequently a second particle accumulation 
sets in for $v \sim (-3.5,-2.8)$ which becomes dominant (second current-reversal). 
The key observation to understand the above-discussed effects is that the velocity regime ($v \sim 2-3$), for which the particle accumulation occurs in case of the first current-reversal
matches the velocity regime of the large elliptic islands of the SPPS (Fig.~\ref{psos}).
The additional above-mentioned particle accumulations occur for velocities which are directly above/below the FISC-velocities of the SPPS. 
These observations indicate, that the structure of the phase-space of the corresponding single-particle system is still of crucial importance for the many-particle dynamics. 
To verify the particle accumulation in certain (regular) parts of the underlying SPPS we lay the PSOS 
(Fig.\ref{psos}) and the position-velocity-distribution of the ensemble for large times ($t/T\sim 2 \cdot 10^5$) on top of each other.
For $\alpha=0$ (Fig.~\ref{overlap} (a)) all particles are situated in the chaotic sea of the SPPS and the regular parts are unoccupied.
The same holds also for the short time behaviour ($t< 10^2 T$) for $\alpha=0.8$ (not shown). Considering the case $\alpha=0.8$ at $t=2\cdot 10^5 T$ 
we observe a vital accumulation of particles in both large elliptic islands centered at $v \approx 2.5$. This accumulation spreads
out into the region of the invariant spanning curves of the SPPS. A further accumulation occurs directly below the lower FISC. 
Since the one in the large islands for positive velocities is dominant, the average velocity has become positive.
We therefore conclude, that both the current-reversal and the high velocity of the DC for $\alpha=0.8$
are caused by a particle accumulation in the regular islands of the corresponding SPPS which are inaccessible for chaotic trajectories in the noninteracting case.
For $\alpha=8.0$, the particle accumulation in the large regular islands takes place for earlier times and it is responsible for the first reversal of the DC.
Fig.~\ref{overlap} (c) shows the corresponding superposition for long times ($t/T \sim 2\cdot 10^5$) and exhibits a dominant accumulation below the lower FISC of the SPPS which illuminates the second current-reversal
with the DC pointing finally in negative direction. 

\begin{figure}[htb]
\begin{center}
\includegraphics[width=0.48\textwidth]{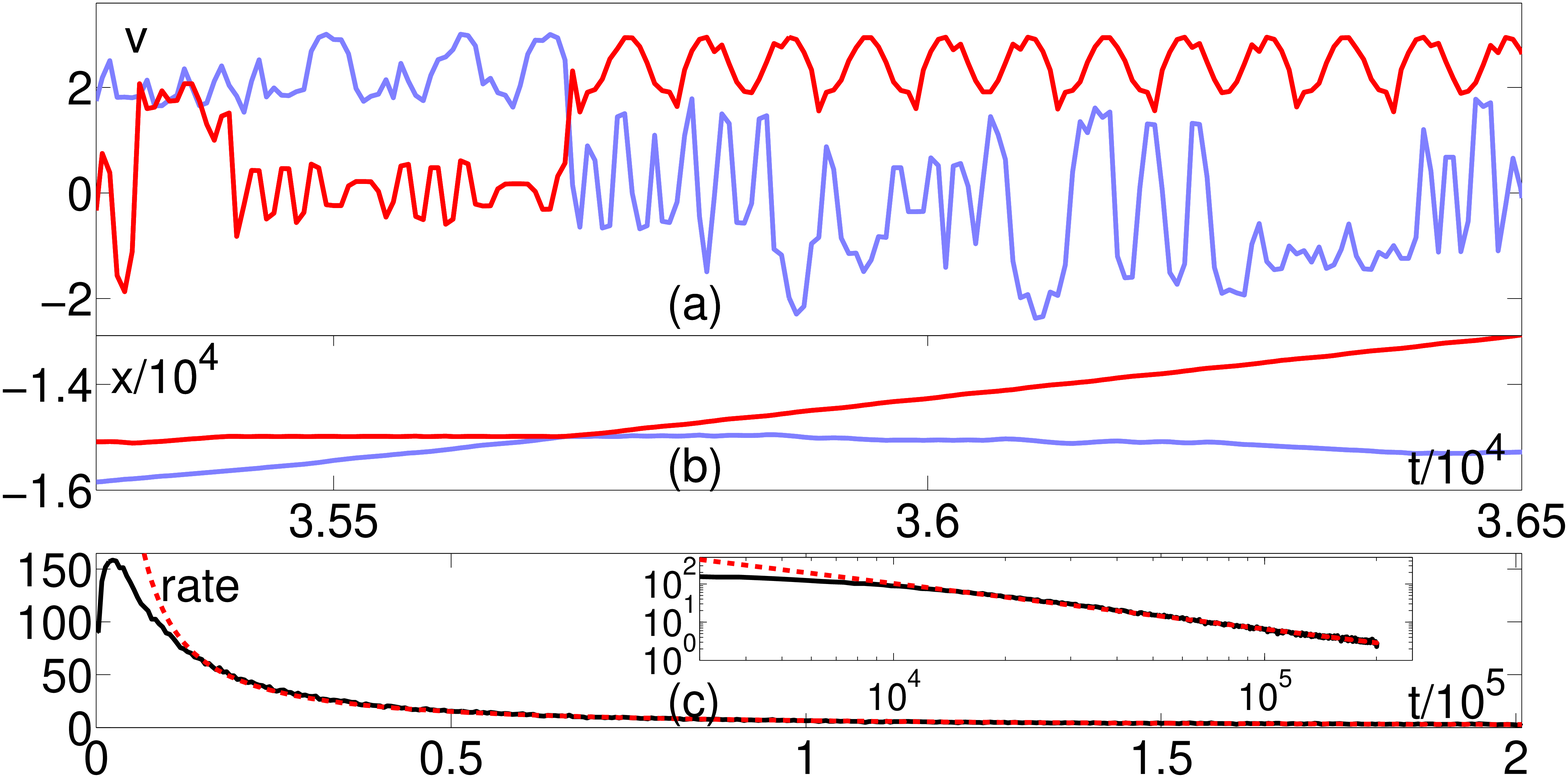}
\caption{The doorway process into a regular island: Extract of the time-evolution of the velocities (a) and positions (b) for $N=2$ and $\alpha=0.8$ in the vicinity of a two-body collision.
(c) Total rate of particle collisions (per time period $T$) for $\alpha=0.8$ (black) and $t^{-1.2}$-fit (red/dashed) with an inset using a logarithmic timescale.}
\label{elem}
\end{center}
\end{figure}
Let us explore the doorway process into the regular parts of the SPPS and the subsequent trapping for long times. 
A typical island-entering event is shown for $N=2,\alpha=0.8$ in Fig.~\ref{elem}. 
Before the two-particles encounter a close collision ($t \sim 3.57\cdot 10^4$ see Fig.~\ref{elem} (b)) their dynamics is chaotic. 
Thereafter one particle is inside the regular island and the other particle exhibits chaotic motion.
The regular motion persists until a further collision occurs: 
then the particle inside the island (red) can either penetrate deeper into the island or it might exit the island and proceed with chaotic motion again.
The interaction-induced doorway to the regular parts of the SPPS is therefore inherently a two-particle process.
Note that contact interactions instead of long-range interactions would only yield the exchange of the velocities of both particles and no island-entering
events would take place.
\paragraph*{Discussion}
It is important to note that the localization onto the elliptic islands
of the SPPS induces a synchronized ballistic particle-barrier motion which
is stable with respect to small perturbations. 
This mechanism combined with the tunability of the regular structures of the
SPPS and their possible change in time via external parameters \cite{petri10,liebchen11,wulff12} provides the perspective of controlling the time-evolution
of the direction and magnitude of DCs.
\\Let us address the question of the balance of entrance and exit processes concerning the regular parts of the SPPS 
in order to illuminate the particle accumulation process.
For a dilute system the mean interaction energy is small
in comparison with the mean kinetic energy of the particles. 
As long as the particles are largely separated in configuration space their Coulomb interaction provides only a weak perturbation
to the single-particle dynamics resulting in general only in a weak perturbation of the regular motion. 
For the case of two particle collisions however, their strong
interaction destroys the tori of the SPPS for a short transient time. 
A random perturbation could in principle change the single-particle dynamics with equal probability for into and out-of
island processes. 
The crucial difference is, that a chaotic trajectory can approach in line with the single-particle dynamics the border of any elliptic island embedded in the chaotic sea, but particles
confined to orbits inside elliptic islands possess a minimal nonzero distance to this border in phase space. 
Consequently, small perturbations can be sufficient to bring a particle across the border inside an elliptic island, but
in order for a particle which is deep inside an elliptic island of the SPPS to leave it,
a whole sequence of adjusted perturbations is necessary. 
For this reason, in the average more particles enter the regular parts of the SPPS than are leaving them.
We have confirmed this by additional simulations for weaker interactions $\alpha=0.08$ for which also the smaller regular islands of the SPPS possess an enhanced occupation.
Since the average distance between the repulsively interacting particles increases with increasing time, the collisional interaction-energy decreases and the doorway process
to the islands becomes less efficient which is in line
with the decrease of the acceleration of the DC observed in Fig.~\ref{means} (d). 
Fig.~\ref{elem} (c) shows that the collisional rate increases for short times but thereafter
decreases rapidly, reaching approximately one collisional event per $10^5$ particles and period $T$ at $t=2\cdot 10^5$. 
For long times the decrease is well described by a $t^{-1.2}$-fit, i.e. it is slightly faster than linear.
The latter leads to a stabilization of the particle accumulation for very long time scales, probably also in the asymptotic regime.
Let us finally briefly discuss the timescales of the dynamics. 
The ergodic filling of the SPPS ($t \ll 10^2$, Fig.~\ref{tev}), current-reversals ($t \sim 10^3 - 2\cdot 10^4$, Fig.~\ref{means})
and the decay of the collisional rate ($t \sim 3 \cdot 10^4$ Fig.~\ref{elem} (c)) happen at separate timescales.
As the incoming particle current into a regular region of the SPPS notably increases with the surface of the latter (i.e. the corresponding particle density increases with the fraction surface/volume)
we obtain particle accumulations first for the small islands, followed by the large islands and then beyond the FISC (Fig.~\ref{tev} (b),(c)).
This current increases with $\alpha$ and
the population of the small and the large islands (Fig.~\ref{tev} (b),(c)) as well as the first current-reversal happen earlier for $\alpha=8.0$ compared to $\alpha=0.8$.
We note that the described mechanism applies to other long-range interactions and in particular even to short-range interactions, but with a significantly lower efficiency.
Clearly, it is possible to design the current-reversals via the structure of the underlying SPPS and the interaction strength.
\\The predicted current-reversals, enhanced ratchet efficiency and the induced particle synchronization should be observable in experiments for electrons in AC-voltage or laser driven arrays of semiconductor heterostructures \cite{linke99}
or cold ions trapped in time-driven lattice potentials \cite{schiavoni03,gommers05,salger09}.

\acknowledgments
We thank C. Petri for fruitful discussions.

\end{document}